\documentclass{article}
\usepackage{amssymb}
\usepackage[centertags]{amsmath}

\begin{document}

\title{Comparison of a general series expansion method
and  the homotopy analysis method}

\author{ Cheng-shi Liu*\\Department of Mathematics\\Northeast Petroleum University\\Daqing 163318, China
\\ *Email: chengshiliu-68@126.com\\ \\Y Liu\\
College of Petroleum Engineering\\Daqing Petroleum Institute\\Daqing
163318, China}

 \maketitle

\begin{abstract}
A simple analytic tool namely the general series expansion method is
proposed to find the solutions for nonlinear differential equations.
By choosing a set of suitable basis functions
$\{e_n(t,t_0)\}_{n=0}^{+\infty}$ such that the solution to the
equation can be expressed by
$u(t)=\sum_{n=0}^{+\infty}c_ne_n(t,t_0)$. In general, $t_0$ can
control and adjust the convergence region of the series solution
such that our method has the same effect as the homotopy analysis
method proposed by Liao, but our method is more simple and clear. As
a result, we show that the secret parameter $h$ in the homotopy
analysis methods can be explained by using our parameter $t_0$.
Therefore, our method reveals a key secret in the homotopy analysis
method.  For the purpose of comparison with the homotopy analysis
method, a typical example is studied in detail.


\end{abstract}

\section{Introduction}

For a given nonlinear differential equation with initial and
boundary conditions, it is well known that we can use the Taylor
series expansion at the initial point to find its solution. On the
other hand, according to the initial and boundary conditions, if we
choose a set of suitable basis functions
$\{e_n(t,t_0)\}_{n=0}^{+\infty}$ to expand the solution, we can
obtain more efficient solution. For example, for a vibrate problem,
we can choose periodic functions such as sine or cosine functions as
the basis functions. Fourier series or transformation method is a
typical example. In general, the convergence region of the series
solution is relative small. How to enlarge the convergence region of
the solution? The homotopy analysis method proposed by Liao[1]
provides a way to solve the problem. We describe the Liao's method
as follows. Consider the nonlinear differential equation
\begin{equation}
Lu+N(u)=0,
\end{equation}
with some initial or boundary conditions, where $L$ is a linear
operator, and $N$ is a nonlinear operator. The key step is to
construct the following homotopy
\begin{equation}
(1-p)[Lu-Lu_0]+ph[Lu+N(u)]=0.
\end{equation}
When $p$ changes from $0$ to $1$, the solution of Eq.(2) changes
from $u_0$ to the solution of Eq.(1). Concretely, take the solution
of Eq.(2) in the following series form
\begin{equation}
u(t,p)=\sum_{n=0}^{+\infty}u_n(t)p^n,
\end{equation}
where $u_n(t)$ are unknown functions to be determined, for $n=0,
1,\cdots.$ Substituting $u(t,p)$ into Eq.(2) and setting the
coefficients of $p^n$ to zero yields the system of the differential
equations of $u_n(t)'s$. Solving the system gives every $u_n$. In
finally, taking $p=1$ gives the solution $u(t)=u(t,1)$ of Eq.(1). In
a series of papers[2-10] and the book[11], liao developed and
applied his method to deal with  a lot of nonlinear problems.
According to different choice of $u_0(t)$, the final form of
solution of Eq.(2) is given by
\begin{equation}
u(t)=\lim_{m\rightarrow \infty}\sum_{n=0}^{m}\mu_{m,n}(h)c_ne_n(t),
\end{equation}
where $ \{e_n(t)\}_{n=0}^{+\infty}$ is a set of functions and the
approaching function $\mu_{m,n}(h)$ satisfies $\lim_{m\rightarrow
\infty}\mu_{m,n}(h)=1$. For example, if we take $e_n(t)=t^n$ for
$n=0,1,2,\cdots$, we obtain the so-called generalized Taylor series
solution. Liao points out that the generalized Taylor series
provides a way to control and adjust the convergence region through
an auxiliary parameter $h$ such that the homotopy analysis method is
particularly suitable for problems with strong nonlinearity[1,11].

It is easy to see that the auxiliary parameter $h$ is the key to the
homotopy analysis method. However, the mathematical meaning of the
parameter $h$ is unknown so that the homotopy analysis method is
still wrapped in a secret veil. This is an open problem in the
theory of the homotopy analysis method. Thus a natural aim is to
solve it.

In the present paper, we propose a general series expansion method
for nonlinear differential equations. By comparing with the homotopy
analysis method, we give the answer of the problem on the
mathematical meaning of $h$ in the homotopy analysis method. In
order to understand the key of my idea, we first prove that the
so-called generalized Newton binomial series at the initial point is
just the usual Newton binomial series expansion at another point
$t_0$. For other basis functions, we also introduce an auxiliary
point $t_0$ whose mathematical meaning is clear. Indeed, we find
that the point $t_0$ can control and adjust the region of the
corresponding series solution. Thus our method includes two key
steps: one is the choice of the basis function, another is the
choice of the point $t_0$. These two simple choices can provide us
with an efficient way to expand the solution and control its
convergence region. By the comparison of these two methods, we
uncovers some secret aspects of the homotopy analysis method. Of
course, as a  simple but efficient method, the general series
expansion method can be directly used to solve nonlinear equations.
We will give other applications in the future.

\section{On the generalized Newton binomial theorem}

In order to understand easily the key of  our idea, we discuss the
 Liao's generalized Newton binomial theorem. For real
number $\alpha (\alpha\neq 0,1,2,3,\cdots)$,   the generalized
Newton binomial theorem is given by
\begin{equation}
(1+t)^{\alpha}=\lim_{m\rightarrow
\infty}\sum_{n=0}^{m}\mu_{\alpha}^{m,n}(h)\binom{\alpha}{n}t^n,
\end{equation}
whose convergence region is
\begin{equation}
-1<t<\frac{2}{|h|}-1, (-2<h<0),
\end{equation}
 where
\begin{equation}
\mu_{\alpha}^{m,n}(h)=(-h)^{n-\alpha}\sum_{j=0}^{m-n}
(-1)^j\binom{\alpha-n}{n}(1+h)^j,
\end{equation}
and
\begin{equation}
\binom{\alpha}{n}=\frac{\alpha(\alpha-1)\cdots(\alpha-n+1)}{n!}.
\end{equation}
The approaching function $\mu_{\alpha}^{m,n}(h)$ satisfies
$\lim_{m\rightarrow \infty}\mu_{m,n}(h)=1$ when $n\geq 1$. Liao[11]
points out that the generalized Newton binomial theorem  provides a
way to control and adjust the convergence region through an
auxiliary parameter $h$.  We next point out that the generalized
Newton binomial theorem is exactly the usual Newton binomial
expansion
 at another point $t_0=-1-\frac{1}{h}$. Indeed, when
$t_0=-1-\frac{1}{h}$, the Taylor expansion at the point $t_0$ is
given by
\begin{eqnarray}
(1+t)^{\alpha}=
(1+t_0)^{\alpha}\sum_{n=0}^{\infty}\binom{\alpha}{n}\frac{1}{(1+t_0)^n}(t-t_0)^n\cr
=\lim_{m\rightarrow
\infty}\sum_{k=0}^{m}\mu_{\alpha}^{m,k}(h)\binom{\alpha}{k}t^k,
\end{eqnarray}
where
\begin{eqnarray}
\mu_{\alpha}^{m,k}(h)=(-h)^{k-\alpha}\sum_{j=0}^{m-k}\binom{\alpha-k}{j}
(h+1)^{j}(-1)^{j}.
\end{eqnarray}
Correspondingly, the convergence region $|t-t_0|<|1+t_0|$ of the
binomial expansion at the point $t_0=-1-\frac{1}{h}$ is just (6).
This result reveals clearly the secret of the auxiliary parameter
$h$ and the essence of the generalized Newton binomial theorem. More
details can be found in Ref.\cite{liu4}. Furthermore, we can prove
that the generalized Taylor series is essentially the usual Taylor
expansion at another point\cite{liu}.

\section{General series expansion method}

For a given differential equation of function $f(t)$
\begin{equation}
N(f,f',\cdots)=0,
\end{equation}
with some original or boundary conditions at the point $t=0$, e.g.,
$f(0)=0, f'(0)=0, f(+\infty)=0$. Firstly, we choose a set of base
functions $e_0(t,t_0),e_1(t,t_0),\cdots$. Then we expand the
solution as a series ,
\begin{equation}
f(t)=\sum_{n=0}^{+\infty}a_ne_n(t,t_0).
\end{equation}
Substituting the series solution into equation (11) and using the
original and boundary conditions, we can determine the values of the
parameters $a_n$ for $n=0,1, \cdots$.

On the other hand, we can expand every $e_n(t,t_0)$ as
\begin{equation}
e_n(t,t_0)=\sum_{m=0}^{+\infty}b_{nm}(t_0)e_m(t), (n=0,1,\cdots),
\end{equation}
where
\begin{equation}
e_m(t)=e_m(t,0).
\end{equation}
Then we have
\begin{equation}
f(t)=\sum_{n=0}^{+\infty}\sum_{m=0}^{+\infty}a_nb_{nm}(t_0)e_m(t),
\end{equation}
Therefore there is a relationship between the point $t_0$ and the
auxiliary parameter $h$ such that
\begin{equation}
f(t)=\lim_{m\rightarrow +\infty}\sum_{n=1}^{m}\mu^{m,n}(h)c_ne_n(t),
\end{equation}
where $c_n$ satisfies
\begin{equation}
f(t)=\sum_{n=1}^{+\infty}c_ne_n(t).\\
\end{equation}

\textbf{Remark 1}. Comparing with the usual series expansion methods
such as power series method whose expanding point is in general the
original point, an important different point of our proposed general
series method is that the expanding point of our series is a motive
point $t_0$ which can control the convergence region.

\textbf{Remark 2}. Ma and Fuchssteiner proposed a powerful approach
for finding exact solutions to nonlinear differential equations
\cite{ma}. The crucial idea is to expand solutions of given
differential equations as functions of solutions of solvable
differential equations, in particular, polynomial and rational
functions. A more systematical theory on decompositions and
transformations is presented very recently in Refs.\cite{ma1} and
\cite{ma2}. The resulting theory unifies many existing approaches to
exact solutions such as the tanh-function methods, the homogeneous
balance method, the exp-function method and the Jacobi elliptic
function method.

\textbf{Remark 3}. The introduction of the parameter $h$ in the
homotopy analysis method is to control the convergence region. In
order to increase the speed of convergence of the series solution,
Liao used the P\'ade approximation to do it(see Ref.[11] for
details). Therefore, we can also use P\'ade approximation to
increase the speed of convergence of the general series solution.

\section{A typical example}

In the chapter 2 of the book [11], Liao studies a typical example
which reads
\begin{equation}
V'(t)=1-V^2(t), \ \ V(0)=0,
\end{equation}
whose exact solution is given by
\begin{equation}
V(t)=\tanh(t).
\end{equation}
Liao constructs the following homotopy
\begin{equation}
(1-q)(\frac{\partial (\Phi(t;q)-V_0(t))}{\partial
t})=qh(\frac{\partial \Phi(t;q)}{\partial t}+\Phi^2(t;q)-1)
\end{equation}
where $q\in [0,1]$. Then we have $\Phi(t;0)=V_0(t), \Phi(t;1)=V(t)$.
We take
\begin{equation}
\Phi(t;q)=V_0(t)+\sum_{m=1}^{+\infty}V_m(t)q^m,
\end{equation}
where $V_m(t)$ are the functions to be determined.  Substituting the
solution into the homotopy equation yields a polynomial of $q$.
Setting the coefficients of the polynomial gives a series of linear
differential equations which solution $V_m(t)$ can be obtained. In
final, we take $q=1$ to give the solution needed. According to the
rule of solution expression, by taking  $V_0(t)=t$,  Liao obtains a
generalized Taylor series solution
\begin{equation}
V(t)=\lim_{m\rightarrow
\infty}\sum_{n=0}^{m}\mu_0^{m,n}(h)\alpha_{2n+1}t^{2n+1},
\end{equation}
where
\begin{equation}
\mu_0^{m,n}(h)=(-h)^n\sum_{k=0}^{m-n}\binom{n-1+k}{k}(1+h)^k,
\end{equation}
and $\alpha_{2n+1}$ are the coefficients of the Taylor expansion of
$\tanh(t)$ at the point $t=0$,
\begin{equation}
\tanh(t)=\sum_{n=0}^{+\infty}\alpha_{2n+1}t^{2n+1},
\end{equation}
whose convergence radius is $\rho_0\approx \frac{3}{2}$ (indeed, we
have $\alpha_{2n+1}=\frac{2^{2k+2}(2^{2k+2}-1)}{(2k+2)!}B_{2k+2}$
where $B_{2k+2}$ are Bernoulli numbers). Liao gives the convergence
region of the generalized Taylor series (22)
\begin{equation}
0\leq t<\rho_0\sqrt{\frac{2}{|h|}-1},
\end{equation}
with $-2<h<0$. In the paper\cite{liu}, we have studied this case in
detail and showed that the generalized Taylor series was just the
usual Taylor series at another point. In the present paper, I study
another set of basis functions. As the same as Liao[11], take the
set of basis functions
\begin{equation}
e_n(t)=\mathrm{e}^{-nt}, (n=0,1,2,\cdots).
\end{equation}
Liao gives the $m$-th order approximate solution
\begin{equation}
V(t)\approx 1+2\sum_{n=1}^m
\{(-1)^n\exp(-2nt)\mu_0^{m,n}(h)\}-\exp(-t)\{1+\frac{h}{2}+\frac{h}{2}\exp(-2t)\}^m.
\end{equation}
When $h=-2$, the solution becomes
\begin{equation}
V(t)\approx 1+2\sum_{n=1}^m
\{(-1)^n\exp(-2nt)\}+(-1)^{m+1}\exp(-(2m+1)),
\end{equation}
which converges to the exact solution in the region $0\leq
t<+\infty$. However, if we expand directly the exact solution using
the basis functions $\{e_n(t)\}_0^{+\infty}$, we have
\begin{equation}
V(t)=1+2\sum_{n=1}^{+\infty} \{(-1)^n\exp(-2nt)\},
\end{equation}
which converges to the exact solution in the region $0< t<+\infty$,
and dose not converge at the point $t=0$. This is a serious
shortcoming for solution (29). Although the Liao's solution (27) can
avoid this weakness, its accurate convergence region can't be given.
Our method can overcome these two difficulties.

We next solve the equation (18) using our general series expansion
method at the point $t_0$ to give the same results with Liao's. We
take the basis functions $e_n(t,t_0)=\{\exp(-2t)-t_0\}^n$ and expand
the solution as
\begin{equation}
V(t)=\sum_{n=1}^{+\infty} a_n\{\exp(-2t)-t_0\}^n,
\end{equation}
and substitute it into Eq.(18) to give
\begin{equation}
a_0=\frac{1-t_0}{1+t_0},  a_{n}=2(-1)^n\frac{1}{(1+t_0)^{n+1}},
n=1,2,\cdots.
\end{equation}
Therefore the series solution is given by
\begin{eqnarray}
V(t)=\frac{1-t_0}{1+t_0}+2\sum_{n=1}^{+\infty}
(-1)^n\frac{1}{(1+t_0)^{n+1}}\{\exp(-2t)-t_0\}^n\cr
=1+2\sum_{n=1}^{+\infty}\{\sum_{k=n}^{+\infty}\binom{k}{n}\frac{t_0^{k-n}}{(1+t_0)^{k+1}}\}(-1)^n\exp(-2nt)\cr
=1+\lim_{m\rightarrow
\infty}2\sum_{n=1}^{m}\{\sum_{k=n}^{m}\binom{k}{n}\frac{t_0^{k-n}}{(1+t_0)^{k+1}}\}(-1)^n\exp(-2nt),
\end{eqnarray}
whose convergence region is
\begin{equation}
-\frac{1}{2}\ln(2t_0+1)<t<+\infty,
\end{equation}
which can be obtained by computing
$|\frac{a_{n+1}}{a_n}(\exp(-2t)-t_0)|<1$ from (30). When $t_0>0$
such as $t_0=1$, the convergence region consists of the region
$0\leq t<+\infty$. Thus $t_0$ can control and adjust the convergence
region. This shows that our method is more simple and efficient in
some degrees than the homotopy analysis method.

Take $t_0=-1-\frac{1}{h}$, and denote
\begin{equation}
\mu^{m,n}(h)=\sum_{k=n}^{m}\binom{k}{n}\frac{t_0^{k-n}}{(1+t_0)^{k+1}}
=(-h)^{n+1}\sum_{k=n}^{m}\binom{k}{n}(h+1)^{k-n}.
\end{equation}
Therefore the solution can be written as
\begin{equation}
V(t)=1+\lim_{m\rightarrow
\infty}2\sum_{n=1}^{m}\mu^{m,n}(h)(-1)^n\exp(-2nt),
\end{equation}
whose convergence region is
\begin{equation}
-\frac{1}{2}\ln(-1-\frac{2}{h})<t<+\infty.
\end{equation}
This result once again shows that the role of our parameter $t_0$ is
the same as the Liao's parameter $h$.

\section{Conclusion}
We propose a general series expansion method to find the solutions
to nonlinear differential equations. Because the expanding point of
the corresponding series solution is a motive point, the method has
a freedom to control the convergence region. Through detailed
analysis of a typical example, we show that we can use the general
series expansion  to give the same result obtained by the homotopy
analysis method. As a result, we reveal the secret of the auxiliary
parameter $h$ which is the key of the homotopy analysis method. In
some degrees our method
is more simple and efficient than the homotopy analysis method.\\

\textbf{Acknowledgements}. Thanks to referees for their valuable
suggestions and calling our attention to existing papers [14] and
[16]. The first author (C S Liu) would like to thank Prof. Wen-Xiu
Ma for his helpful discussion.

\end{document}